\documentclass[aps,prl,twocolumn,showpacs,preprintnumbers,amsmath,amssymb]{revtex4-1}
\usepackage{graphicx}
\usepackage{color}
\usepackage[normalem]{ulem}

\renewcommand\sout{\bgroup \color{red}\ULdepth=-.5ex \ULset}
\newcommand\soutb{\bgroup \color{blue} \ULdepth=-.5ex \ULset}

\usepackage{graphicx}
\usepackage{dcolumn}
\usepackage{bm}

\begin{document}

\preprint{}

\title{
Large long-distance contributions to the electric dipole moments of charged leptons in the standard model
}

\author{Yasuhiro~Yamaguchi$^{1,2}$}
  \affiliation{$^1$Advanced  Science  Research  Center,  Japan  Atomic  Energy  Agency  (JAEA),  Tokai  319-1195,  Japan}
  \affiliation{$^2$RIKEN Nishina Center, RIKEN, 
  Wako, Saitama 351-0198, Japan}
  \email{yamaguchi.yasuhiro@jaea.go.jp}
\author{Nodoka~Yamanaka$^{2,3}$}
  \affiliation{$^2$Amherst Center for Fundamental Interactions, Department of Physics, University of Massachusetts, Amherst, 
Massachusetts 01003, USA}
  \affiliation{$^3$Yukawa Institute for Theoretical Physics, Kyoto University, 
  Kitashirakawa-Oiwake, Kyoto 606-8502, Japan}
  \email{nyamanaka@umass.edu}

\date{\today}

\begin{abstract}
We reevaluate the electric dipole moment (EDM) of charged leptons in the standard model using 
hadron effective models.
We find 
{unexpectedly} 
large EDM generated by the hadron level long-distance effect, 
$d_e = { 5.8 \times 10^{-40} }$,  
$d_\mu = { 1.4 \times 10^{-38}}$, 
and $d_\tau = { -7.3 \times 10^{-38} }$ ecm, 
with an error bar of 70\%, exceeding the conventionally known four-loop level elementary contribution by several orders of magnitude.
\end{abstract}

\pacs{11.30.Er,12.15.Lk,13.20.-v,13.40.Em}
\maketitle

The electric dipole moment (EDM) \cite{edmreview}, 
the linear response of the energy of a system against electric field $H=-{\mathbf d} \cdot {\mathbf E}$, is
often quoted as the most sensitive observable to the CP violation beyond the standard model (SM). It is required 
to explain the baryon number asymmetry of the Universe \cite{Sakharov:1967dj,Farrar:1993hn,Huet:1994jb}, and active search using various systems such as the neutron \cite{Abel:2020gbr}, atoms \cite{Regan:2002ta,Bishof:2016uqx,Graner:2016ses,Sachdeva:2019mun}, molecules \cite{Hudson:2011zz,Kara:2012ay,Baron:2013eja, Cairncross:2017fip, Andreev:2018ayy}, muons \cite{Bennett:2008dy}, or $\tau$ leptons \cite{tauedm}, is currently 
being carried out.

The measurements of the electron EDM using paramagnetic systems are especially attracting attention thanks to 
their relativistic enhancement by the strong internal electric field \cite{EDMrelativistic}, extensively calculated in theoretical works \cite{atomicedmtheory}.
The experimental upper limit of the electron EDM was successively updated since the 
1960s~\cite{Sandars:1964zz,Weisskopf:1968zz,Stein:1969zz,Player:1970zz,Murthy:1989zz,Abdullah:1990nh,Commins:1994gv,Regan:2002ta,Hudson:2011zz,Baron:2013eja,Andreev:2018ayy}, and currently shows a record of 
$|d_e| < 1.1 \times 10^{-29}e$ cm \cite{Andreev:2018ayy}.
There is also much effort to push it down using paramagnetic atoms trapped in three-dimensional optical lattice \cite{Chin:2001zz,Sakemi:2011zz}, polyatomic molecules \cite{Kozyryev:2017cwq,Aggarwal:2018pru}, or polar molecules and inert gas matrix \cite{Vutha:2017pej}, etc.
The experimental studies of the EDM of the muon and the $\tau$ lepton are also established fields, with the former one measured using storage rings \cite{Bennett:2008dy} and the latter one extracted from the precision analysis of collider experimental data \cite{tauedm}.
The measurability of the muon EDM using storage rings was also recently deeply discussed in the context of the general relativity \cite{GREDM}.

\begin{figure}[htb]
\begin{center}
\includegraphics[scale=0.6]{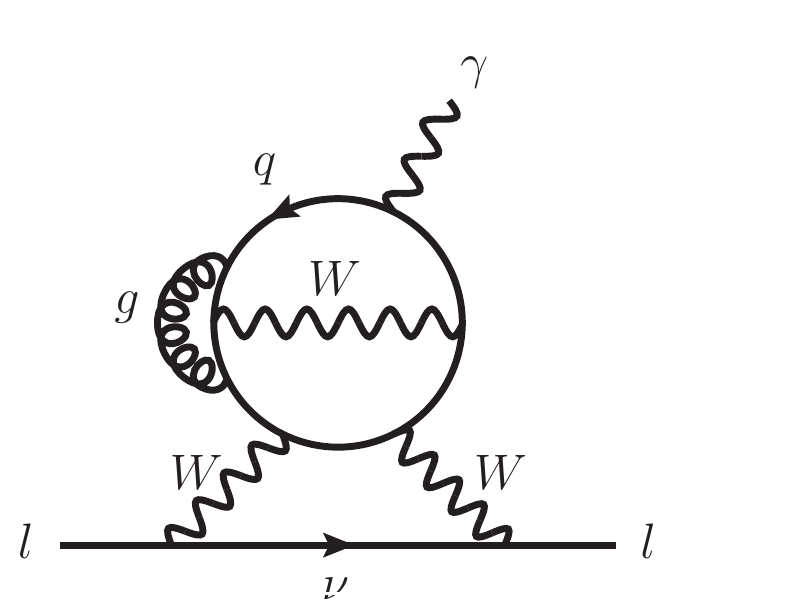}
\caption{
Example of four-loop level diagram contributing to the EDM of charged leptons generated by the CKM matrix at the quark level.
}
\label{fig:electron_EDM_SM_elementary}
\end{center}
\end{figure}

One of the most attractive advantages
of the EDM is that the effect of the complex phase of the Cabibbo-Kobayashi-Maskawa (CKM) matrix \cite{Kobayashi:1973fv}, which is the representative CP violation of the SM, is extremely small, at least for all known systems.
There, the EDMs of light quarks \cite{Shabalin:1978rs,Shabalin:1980tf,Shabalin:1982sg,Eeg:1982qm,Eeg:1983mt,Khriplovich:1985jr,Czarnecki:1997bu} and electrons \cite{Pospelov:1991zt,Booth:1993af,Pospelov:2013sca} appear from the three- and four-loop levels, with the estimated values $d_{u,d} \sim O(10^{-35})$ 
and $d_e \sim O(10^{-45})e$ cm, respectively (for the electron, an example of the four-loop level diagram is displayed in Fig. \ref{fig:electron_EDM_SM_elementary}).
The Weinberg operator (chromo-EDM of gluons) is also very small, yielding an EDM to the neutron of $O(10^{-40})e$ cm \cite{Pospelov:1994uf}.
This extreme suppression is due to the antisymmetry of the Jarlskog invariant \cite{Jarlskog:1985ht} in the exchange of flavor, which is an important consequence of the Glashow-Iliopoulos-Maiani (GIM) mechanism \cite{Glashow:1970gm,Ellis:1976fn,Pospelov:2013sca}, leading to the cancellation of almost equal terms 
and thus bringing additional factors of quark masses. 
It can actually be proven that, under the GIM mechanism, the EDM of the charged lepton evaluated at the elementary level suffers from the suppression factor $m_b^2 m_c^2 m_s^2 / m_t^6$ at all orders of perturbation, yielding at most $d_e \sim 10^{-48} e$ cm \cite{Yamaguchi:2020dsy}. 

\begin{figure}[htb]
\begin{center}
\includegraphics[width=9.2cm]{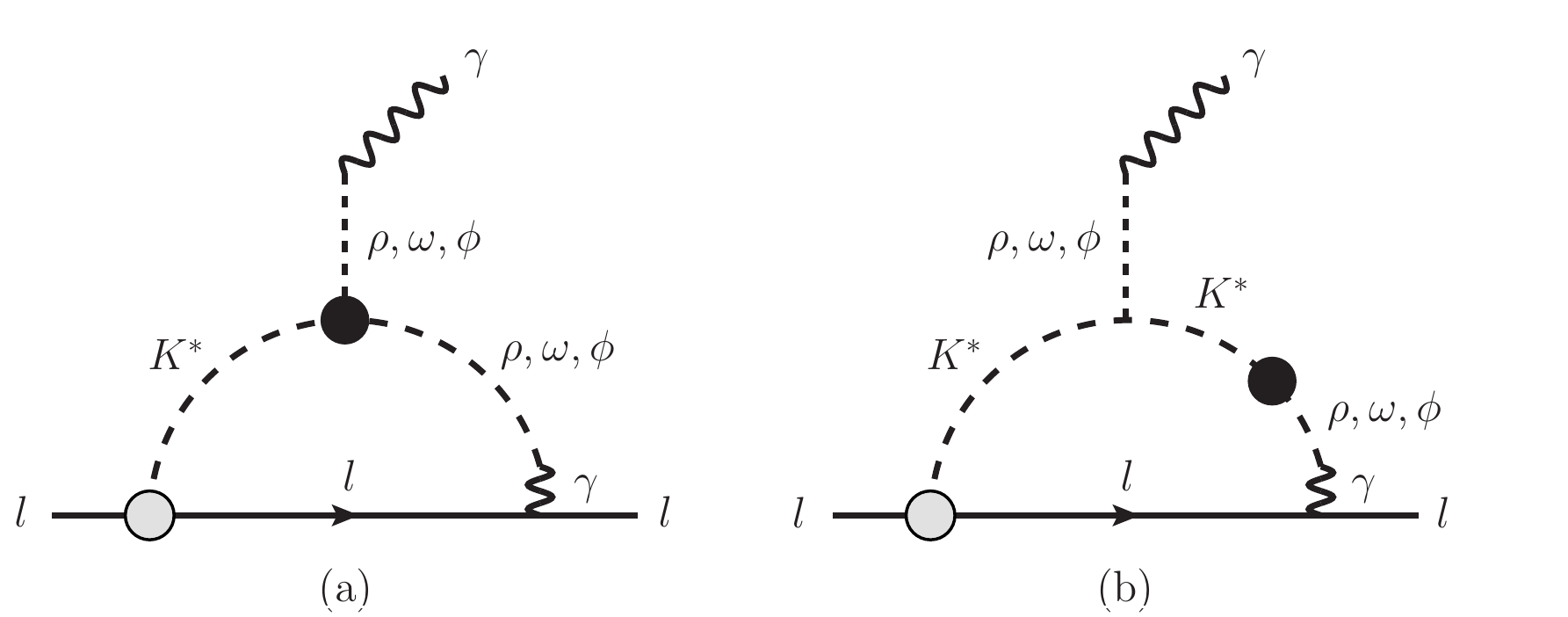}
\caption{
Long-distance contribution to the charged lepton EDM  generated in the SM, with the (a) strong and (b) weak three-vector meson interactions.
Other symmetric diagrams are not displayed.
The $|\Delta S|=1$ 
semileptonic interaction (gray blob) and the $|\Delta S|=1$ hadronic interaction (black blob) are chosen so as to form the Jarlskog invariant.
}
\label{fig:electron_EDM_long-distance}
\end{center}
\end{figure}

On the other hand,
the CKM contributions to the EDM of composite systems are believed to be more enhanced thanks to the {\it long-distance effect}, where the Jarlskog combination is realized with two distinct $|\Delta S|=1$ hadron level interactions.
As for the nucleon EDM, this contribution is larger than the quark EDM and chromo-EDM contributions by 
2 or 3 orders of magnitude \cite{Khriplovich:1981ca,McKellar:1987tf,Seng:2014lea}.
In the case of nuclear EDM or the nuclear Schiff moment, it was also shown that the long-distance effect is also much larger than the short-distance one \cite{Yamanaka:2015ncb,Yamanaka:2016fjj,Lee:2018flm}.
Then, what about the lepton EDM?
We actually found that a similar long-distance mechanism also happens for the case of the EDM of charged leptons.
At the hadronic level, 
one-loop level diagrams (see Fig. \ref{fig:electron_EDM_long-distance}) give the leading contribution
and the cancellation is much milder because the loop momenta, given by the hadron masses, are sufficiently different between diagrams, so that we expect a much larger EDM.
In this Letter, we report on the 
evaluation
of this new contribution in the hadronic effective model.

The leading order contribution of the CKM matrix to the lepton EDM is constructed with at least two $W$ boson exchanges. 
To avoid severe GIM cancellation, we have to split the short-distance flavor changing process at least into two parts, while keeping the Jarlskog combination of the CKM matrix elements.
This condition means that the largest contribution to the lepton EDM is generated at the hadron level (the long-distance effect).
The largest long-distance contribution should involve unflavored and $|S|=1$ mesons rather than open heavy flavored ones ($c,b$), which are much heavier.
Another important condition is that the charged lepton EDM is generated by a one-loop level diagram involving vector mesons, because the interaction of pseudoscalar mesons with the lepton will change the chirality, suppressing the EDM by at least by a factor of $m_l^2$ ($l=e, \mu , \tau$).
The charged lepton EDM is then generated by radiative corrections involving a $K^*$ meson.
The one-loop level diagram must not have a neutrino in the intermediate state of the long-distance process, since the chirality flip is small and the charge is neutral.
The $K^*$ meson must therefore change to an unflavored meson, which in turn becomes a photon 
that will be absorbed by the charged lepton.
Under such conditions, we may draw diagrams 
as shown in Fig.~\ref{fig:electron_EDM_long-distance}.

Let us now introduce the interactions to calculate the diagrams of Fig. \ref{fig:electron_EDM_long-distance}.
It is convenient to describe the $|\Delta S|=0$ vector meson interactions with the hidden local symmetry (HLS) formulation \cite{HLS}.
The HLS is a framework introduced to extend the domain of applicability of chiral perturbation to include vector meson resonances, and it is successful in phenomenology.
It generates three-vector meson interactions as follows:
\begin{eqnarray}
{\cal L}_{\rm HLS}^{VVV}
&=&
ig {\rm tr} [(\partial_\mu V_\nu - \partial_\nu V_\mu) V^\mu V^\nu]
,
\label{eq:HLS}
\end{eqnarray}
where 
$
g\equiv \frac{m_\rho}{2f_\pi} \sim 4.2
$, and 
\begin{eqnarray}
V^\mu \equiv
\left(
\begin{array}{ccc}
(\rho^0+\omega ) /\sqrt{2} & \rho^+ & K^{*+} \\
\rho^- & (-\rho^0+\omega ) /\sqrt{2} & K^{*0} \\
K^{*-} & \bar K^{*0} & \phi \\
\end{array}
\right)^\mu.
\end{eqnarray}
We note that this Lagrangian is renormalizable if we assume the vector meson mass has been generated by the Higgs mechanism.

\begin{figure}[htb]
\begin{center}
\includegraphics[width=9cm]{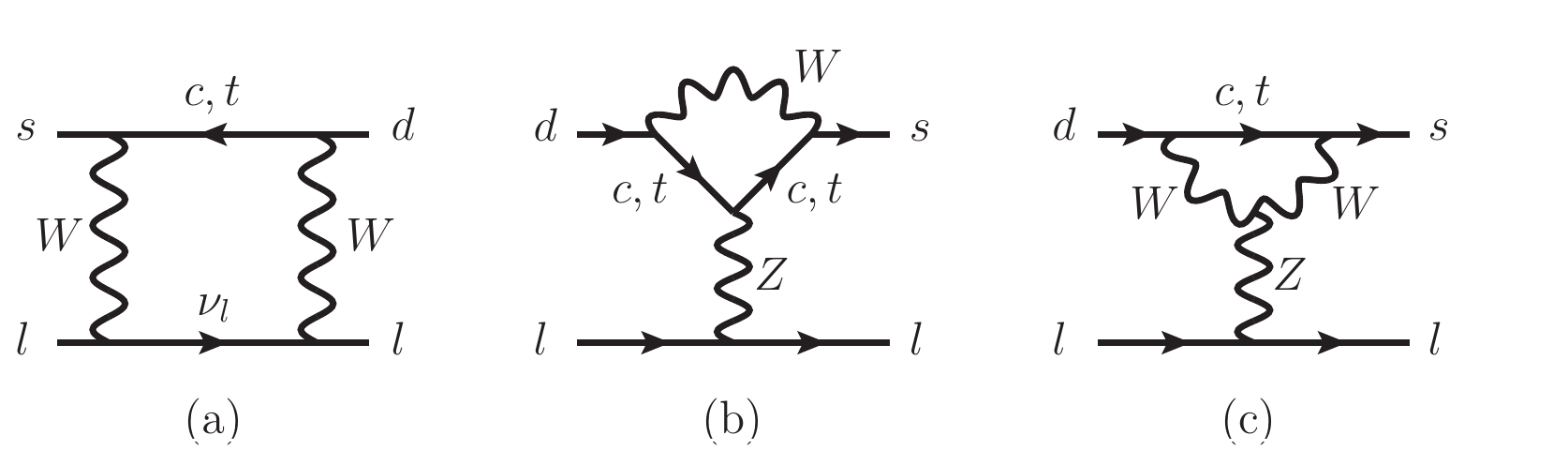}
\caption{
Short-distance contribution to the $K^*$-charged lepton interaction, with $l=e,\mu , \tau$.
The diagrams (a), (b) and (c) are the contributions of the box diagram with the two $W$ exchanges and the penguin diagrams with the quark-$Z$ and $W$-$Z$ vertices, respectively.
}
\label{fig:semileptonic_one-loop}
\end{center}
\end{figure}

Let us now model the weak interaction at the hadron level.
From Fig. \ref{fig:electron_EDM_long-distance}, the $|\Delta S|=1$ weak interaction appears in the semileptonic creation of $K^*$ and 
in the transition between $K^*$ and chargeless vector mesons $\rho , \omega , \phi$.
In Fig. \ref{fig:electron_EDM_long-distance}, the $K^*$-lepton vertex does not change the charge of the lepton, so this interaction must effectively be generated by a loop involving 
the $W$ boson so as to change twice the quark flavor at short distance~\cite{Buchalla:1995vs}, as shown in Fig. \ref{fig:semileptonic_one-loop}.
Then it is best to also include heavy flavored quarks in this loop as well to derive benefit from the large loop momentum if we wish to maximize the effective coupling.
Therefore, this $|\Delta S|=1$ effective interaction is attributed the CKM matrix elements $V_{cs}V_{cd}^*$ or $V_{ts}V_{td}^*$.
It also has to be parity violating, otherwise the $|\Delta S|=1$ meson transition has to create axial vector mesons, which are heavier.

The parity violating interaction between $K^*$ and the charged lepton is given by
\begin{equation}
{\cal L}_{K^*ll}
=
g_{K^*ll} 
K^{*}_\mu
\bar l \gamma^\mu \gamma_5 l
+{\rm (H.c.)}
,
\end{equation}
where $K^*_\mu$ is the field operators of the $K^*$ meson.
In the limit of zero momentum exchange, the coupling constant is given by 
\begin{equation}
{\rm Im} ( g_{K^*ll} ) \epsilon^{K^{*}}_\mu
=
{\rm Im} ( V_{ts}^*V_{td} )
\langle 0 | \bar s \gamma_\mu d | K^* \rangle 
I_{dsll}
, 
\label{eq:K*eecoupling}
\end{equation}
where we fixed the complex phases of $V_{ud}V_{us}^*$ to be real.
The $K^*$ meson matrix element is defined by $\langle 0 | \bar s \gamma_\mu d | K^* \rangle = m_{K^*} f_{K^*} \epsilon^{K^{*}}_\mu $, where $\epsilon^{K^{*}}_\mu $, $m_{K^*}=890$ and 
$f_{K^*}= 204$ MeV 
\cite{decayconstant} are the polarization vector, the mass, and the phenomenologically derived decay constant of $K^*$, respectively. 
The quark level amplitude $I_{dsll}$ can be obtained by calculating the one-loop level diagrams of Fig. \ref{fig:semileptonic_one-loop}.
By neglecting all external momenta [which are $O(\Lambda_{\rm QCD})$] and imposing $m_t, m_W \gg m_c$, we have
\begin{equation}
I_{dsll}
=
{ 3.2}
\times 10^{-8}\, {\rm GeV}^{-2}
.
\end{equation}
This value is quite consistent in absolute value with that of the naive dimensional analysis 
{
$I_{dsll} \sim \frac{\alpha_{\rm QED}^2}{4\sin^4 \theta_W m_W^2} \sim 4.3 \times 10^{-8}\, {\rm GeV}^{-2}$.
}

We now model the $K^*$-$V$ transition ($V=\rho , \omega , \phi$), which is either a two-point vertex or a three-point one.
It is generated by the $|\Delta S| =1$ four-quark effective Hamiltonian  
\begin{equation}
{\cal H}_{eff} (\mu)
=
\frac{G_F}{\sqrt{2}} V_{us}^* V_{ud}
\sum_{i=1}^6
 z_i (\mu) Q_i (\mu)
+ {\rm H.c.}
,
\label{eq:effhamibelowmc}
\end{equation}
where the Fermi constant $G_F = 1.166\,37 \times 10^{-5} {\rm GeV}^{-2}$.
Here $Q_i$ ($i=1 \sim 6$) are defined as \cite{buras} 
\begin{eqnarray}
Q_1 
&\equiv&
\bar s_\alpha \gamma^\mu (1-\gamma_5) u_\beta \cdot \bar u_\beta \gamma_\mu (1-\gamma_5) d_\alpha
,
\label{eq:q1}
\\
Q_2 
&\equiv&
\bar s_\alpha \gamma^\mu (1-\gamma_5) u_\alpha \cdot \bar u_\beta \gamma_\mu (1-\gamma_5) d_\beta
,
\label{eq:q2}
\\
Q_3 
&\equiv&
\bar s_\alpha \gamma^\mu (1-\gamma_5) d_\alpha \cdot \sum_{q=u,d,s} \bar q_\beta \gamma_\mu (1-\gamma_5) q_\beta
,
\label{eq:q3}
\\
Q_4
&\equiv&
\bar s_\alpha \gamma^\mu (1-\gamma_5) d_\beta \cdot \sum_{q=u,d,s} \bar q_\beta \gamma_\mu (1-\gamma_5) q_\alpha
,
\label{eq:q4}
\\
Q_5
&\equiv&
\bar s_\alpha \gamma^\mu (1-\gamma_5) d_\alpha \cdot \sum_{q=u,d,s} \bar q_\beta \gamma_\mu (1+\gamma_5) q_\beta
,
\label{eq:q5}
\\
Q_6
&\equiv&
\bar s_\alpha \gamma^\mu (1-\gamma_5) d_\beta \cdot \sum_{q=u,d,s} \bar q_\beta \gamma_\mu (1+\gamma_5) q_\alpha
,
\label{eq:q6}
\end{eqnarray}
where $\alpha$ and $\beta$ are the color indices.
For the case of the $|\Delta S| =1$ four-quark interaction, the renormalization from the electroweak scale $\mu = m_W$ to the hadronic scale $\mu = 1$ GeV changes the Wilson coefficients $z_i$.
From the numerical calculation of the running in the next-to-leading logarithmic approximation \cite{buras,Yamanaka:2015ncb}, we obtain
\begin{equation}
{\bf z} (\mu = 1 \, {\rm GeV})
=
\left(
\begin{array}{c}
-0.107 \cr
1.02 \cr
1.76 \times 10^{-5} \cr
-1.39 \times 10^{-2}  \cr
6.37 \times 10^{-3} \cr
-3.45 \times 10^{-3} \cr
\end{array}
\right)
.
\end{equation}

\begin{figure}[htb]
\begin{center}
\includegraphics[width=9cm]{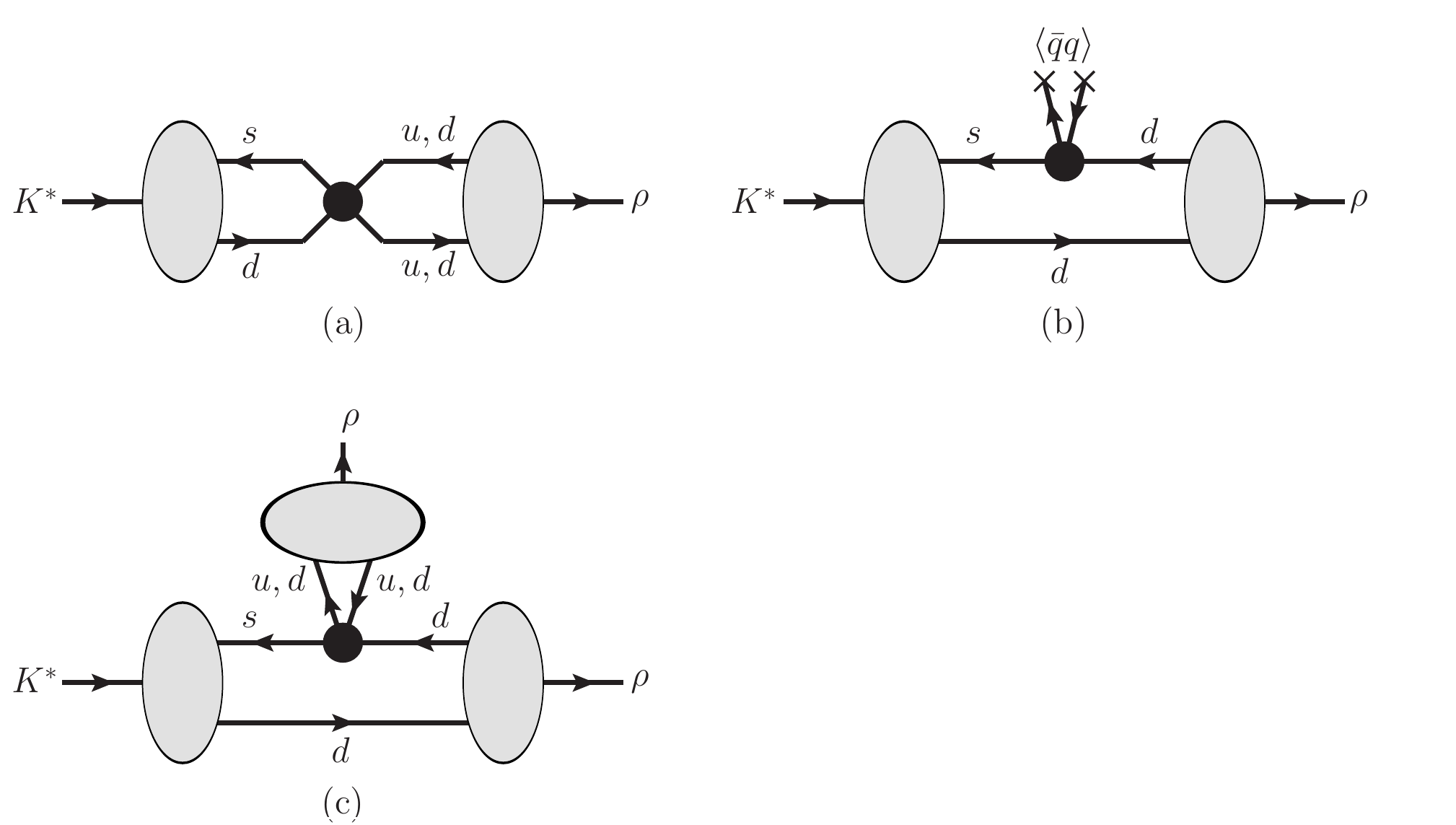}
\caption{
Factorization of the $|\Delta S|=1$ vector meson vertices 
($|\Delta S|=1$ meson transition), with 
the (a) two-quark process, (b) one-quark process, and (c) three-meson interaction.
The double crosses with ``$\langle \bar q q \rangle$'' denote the chiral condensate $\langle 0 | \bar q q |0 \rangle$ ($q=d,s$).
The black blob denotes the $|\Delta S| =1$ four-quark interaction.
There are similar diagrams with the $\rho$ meson replaced by $\omega$ or $\phi$ mesons.
}
\label{fig:factorization}
\end{center}
\end{figure}

We use the standard factorization to derive the $|\Delta S| =1$ vector meson interaction from the $|\Delta S| =1$ four-quark interaction of Eq. (\ref{eq:effhamibelowmc}). 
We first construct the $|\Delta S| =1$ meson transition in the factorization with vacuum saturation approximation \cite{Lee:1972px,Shrock:1978dm}.
The Lagrangian of the weak $|\Delta S| =1$ vector meson transition is given by
\begin{eqnarray}
{\cal L}_{V K^*}
&=&
V_{ud}V_{us}^*
\hspace{-0.5em}
\sum_{V = \rho , \omega , \phi}
\hspace{-0.5em}
g_{ V  K^*} {V}^\nu K^*_\nu
+{\rm H.c.}
,
\label{eq:meson-transition}
\end{eqnarray}
where $V^\nu$ is the field operator of the $\rho_0$, $\omega$, or $\phi$ mesons.
The $|\Delta S| =1$ four-quark interaction has two distinct contributions, as shown in 
Figs.~\ref{fig:factorization}(a) and \ref{fig:factorization}(b).
The former one is generated by all $Q_i$'s, while the latter is only possible through the Fierz transform of $Q_5$ and $Q_6$.
The couplings can be calculated in the factorization as
\begin{eqnarray}
\langle \rho | \bar s \gamma^\mu d \, \bar q \gamma_\mu q | K^* \rangle
&\approx&
\langle 0 | \bar s \gamma^\mu d | K^* \rangle \langle \rho | \bar q \gamma_\mu q | 0 \rangle
,
\label{eq:vacuumsaturation1}
\\
\langle \rho | \bar s d \, \bar d d | K^* \rangle
&\approx&
\langle \rho | \bar s d | K^* \rangle \langle 0 | \bar d d | 0 \rangle
,
\label{eq:vacuumsaturation2}
\end{eqnarray}
where $q=u,d,s$.
We note that the vacuum saturation approximation gives the leading contribution in the large $N_c$ expansion for mesonic processes. 
As shown below Eq.~(\ref{eq:K*eecoupling}), the vector meson matrix elements are related to the decay constants \cite{decayconstant}.
The chiral condensate $\langle 0 | \bar qq | 0 \rangle$ is derived from the Gell-Mann-Oakes-Renner relation, and the vector meson scalar densities
 $\langle \rho^0 |\bar{s}d|K^{\ast 0}\rangle =-1.14$, 
 $\langle \omega |\bar{s}d|K^{\ast 0}\rangle=1.88$,  and
 $\langle \phi |\bar{d}s|K^{\ast 0}\rangle=2.14$ GeV
are extracted from the lattice QCD result of the quark mass dependence of the vector meson mass \cite{Leinweber:2001ac,Guo:2018zvl} and flavor $SU(3)$ symmetry.
The appropriate scale of the factorization procedure was chosen as $\mu = 1$ GeV.
Using these input parameters, the coupling constants are given by 
$g_{\rho K^*} = 4.4 \times 10^{-8} $, 
$g_{\omega K^*} = 3.4 \times 10^{-8} $, and 
$g_{\phi K^*} = -6.6 \times 10^{-9} {\rm GeV}^2$.

Let us also construct the weak three-meson interactions.
Its Lagrangian is given by
\begin{eqnarray}
&&
{\cal L}^V_{V' K^*}
=
V_{ud} V_{us}^*
\hspace{-1em}
\sum_{V,V'=\rho , \omega , \phi}
\hspace{-1em}
g^V_{ V'  K^*} V_\mu {V'}^\nu i \overleftrightarrow \partial^\mu K^*_\nu
+{\rm H.c.}
, \ \ \ 
\label{eq:three-meson}
\end{eqnarray}
where $A \overleftrightarrow \partial^\mu B \equiv A (\partial^\mu B) - (\partial^\mu A) B$.
Again by using the vacuum saturation approximation, we have
\begin{eqnarray}
\langle \rho \, | \bar q \gamma_\mu q\, \bar s \gamma^\mu d |K^* \rho \rangle
&\approx&
\langle 0 | \bar q \gamma_\mu q | \rho \rangle \langle \rho | \bar s \gamma^\mu d | K^* \rangle
.
\ \ \ \ 
\label{eq:vacuumsaturation3}
\end{eqnarray}
where $\langle V (p') | \bar s \gamma^\mu d | K^* (p) \rangle \approx  -(p^\mu + p'^\mu)\epsilon^{(V)*}_\nu \epsilon^{(K^*)\nu}$ ($V= \rho , \omega , \phi$).
The coupling constants are then given by $g^\rho_{\rho K^*} = -g^\rho_{\omega K^*} = -g^\rho_{\phi K^*}/\sqrt{2} = 1.7 \times 10^{-7}$, $g^\omega_{\rho K^*} = -g^\omega_{\omega K^*} = -g^\omega_{\phi K^*}/\sqrt{2} = 1.4 \times 10^{-7}$, 
and $g^\phi_{\rho K^*} = -g^\phi_{\omega K^*} = -g^\phi_{\phi K^*}\sqrt{2} = -1.8 \times 10^{-8}$.

After evaluation, we obtain the following lepton EDM:
\begin{eqnarray}
d_e^{\rm (SM)} 
&=&
{ 5.8 \times 10^{-40} }
e \, {\rm cm}
,
\\
d_\mu^{\rm (SM)} 
&=&
{ 1.4 \times 10^{-38}}
e \, {\rm cm}
,
\\
d_\tau^{\rm (SM)} 
&=&
{ -7.3 \times 10^{-38} }
e \, {\rm cm}
.
\end{eqnarray}
These values are much larger than the estimation at the four-loop level ($d_e^{\rm (SM)} \sim 10^{-48}e$ cm) \cite{Pospelov:1991zt,Booth:1993af,Pospelov:2013sca,Yamaguchi:2020dsy}.
As anticipated in the beginning of this Letter, the enhancement is due to the absence of severe GIM 
cancellation (antisymmetry in the interchange of quark flavor \cite{Pospelov:2013sca}) thanks to the scatter of the typical loop momentum according to the hadron masses, whereas in the short-distance case, the momenta were made almost equal by the heavy top quark or $W$ boson.
This is a general feature of the long-distance effect, and similar enhancement could be seen in the case of the neutron EDM compared to the short-distance quark EDM \cite{Czarnecki:1997bu,McKellar:1987tf,Seng:2014lea,Pospelov:2013sca}.

We also analyze the potential systematics of our study.
The first large source of uncertainty is that due to the renormalization of the $|\Delta S|=1$ four-quark operators.
It may be estimated by looking at the variation of Wilson coefficients between $\mu = 0.6$ and $\mu = m_c = 1.27$ GeV, yielding about 10\%.
The second important source of systematics is the factorization used to calculate the vector meson interactions.
As we saw above, the vacuum saturation approximation gives the leading order effect in the large $N_c$, so we expect the error bar to be about $1/N_c \sim O(40\%)$.
Finally, we have to mention the hadronic uncertainty, associated with 
higher order contribution of HLS. This has been estimated to be $\sim O(50\%)$ as discussed in Ref.~\cite{HLS}.
In all, the theoretical uncertainty should be $\sim 70\%$.
To further quantify this hadronic contribution, lattice QCD techniques used in the evaluations of the light-by-light scattering~\cite{Blum:2019ugy} and $K\rightarrow \pi\pi$ decay~\cite{Bai:2015nea} may play an important role.

\begin{figure}[htb]
\begin{center}
\includegraphics[width=8.5cm]{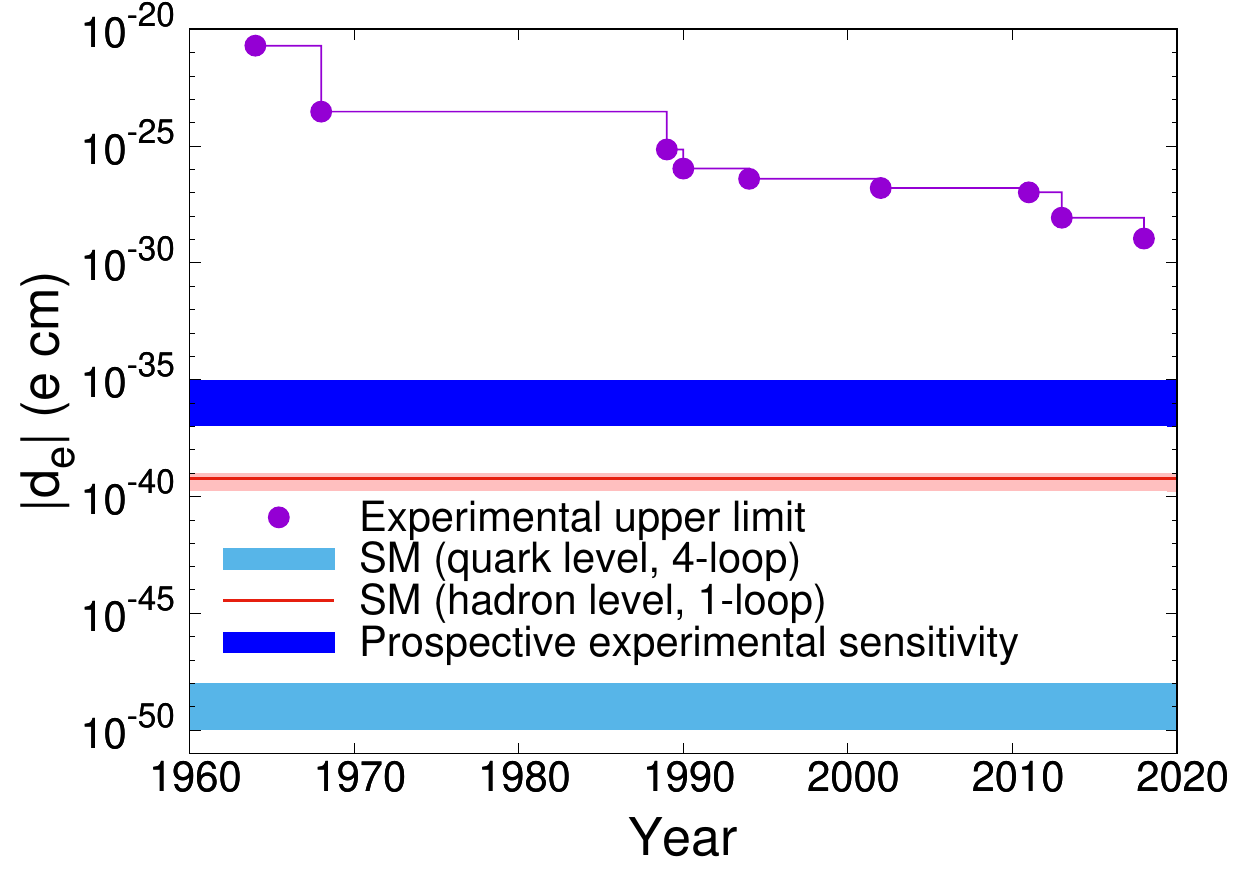}
\caption{
Plot of the SM predictions (quark and hadron levels) of the electron EDM compared with the experimental upper limits updated from the 
1960s
\cite{Sandars:1964zz,Weisskopf:1968zz,Player:1970zz,Murthy:1989zz,Abdullah:1990nh,Commins:1994gv,Regan:2002ta,Hudson:2011zz,Baron:2013eja,Andreev:2018ayy}.
The prospective experimental sensitivity \cite{Vutha:2017pej} is also shown.
}
\label{fig:electronEDM_record}
\end{center}
\end{figure}

In conclusion, we evaluated for the first time the hadronic level contribution to the EDM of charged leptons.
As a result, we found that this long-distance effect is much larger than the previously known one, which was estimated at the elementary level.
The main reason of the enhancement at the hadronic level is because we could avoid severe suppression due to the GIM mechanism.
In Fig. \ref{fig:electronEDM_record}, we plot the EDM of the electron in the SM compared with the progress of the experimental accuracy.
The electron EDM obtained in this 
Letter work is  $d_e \sim 10^{-39}e$ cm, 
which is still well below the current sensitivity of the molecular beam experiments \cite{Andreev:2018ayy}.
The EDM experiments are, however, improving very fast, and we have to be very sensitive to their progress and to proposals with new ideas, with some of them claiming to be able to ideally reach the level of $O(10^{-35}-10^{-37})e$ cm in the {\em statistical} sensitivity~\cite{Vutha:2017pej}.
This potential breakthrough, combined with our result, is maybe cautioning us that we have to be careful with the statement that the window left for the electron EDM to new physics beyond the SM is almost infinite.
We also note that, in EDM experiments using atoms and molecules, the CP-odd electron-nucleon interaction may also contribute at the same order 
{ \cite{Pospelov:2013sca,PhysRevA.93.062503}}.
To truly distinguish the effect of the electron EDM, other independent tests are also required.

We also point out that this long-distance effect is also generated by other semileptonic and quark flavor violating processes.
The most interesting target should be the physics of the $B$ meson decay anomaly, recently suggested by several $B$ factory experiments
\cite{Hiller:2003js,Bfactory} or the recent result of $K$ meson decay of KOTO experiment \cite{KOTOresult,Kitahara:2019lws}.

\begin{acknowledgments}
The authors thank J. de Vries and T. Kugo for useful discussions and comments.
They also thank T. Morozumi for pointing out the contribution of the semileptonic penguin diagram.
A part of the numerical computation in this work was carried out at the Yukawa Institute Computer Facility.
This  work is  supported  in  part  by  the  Special  Postdoctoral
Researcher Program  (SPDR) of RIKEN (Y.Y.).
\end{acknowledgments}


\begin{thebibliography}{999}

\bibitem{edmreview}
X.~G.~He, B.~H.~J.~McKellar, and S.~Pakvasa, 
Int.\ J.\ Mod.\ Phys.\ A {\bf 04}, 5011 (1989); {\bf 06}, 1063(E) (1991);
W.~Bernreuther and M.~Suzuki, 
Rev.\ Mod.\ Phys.\  {\bf 63}, 313 (1991); {\bf 64}, 633(E) (1992); 
S.~M.~Barr, 
Int.\ J.\ Mod.\ Phys.\ A {\bf 08}, 209 (1993);
%
I.~B.~Khriplovich and S.~K.~Lamoreaux, 
{\it CP Violation without Strangeness: Electric Dipole Moments of Particles, Atoms, and Molecules} (Springer, Berlin, 1997), p. 230;
J.~S.~M.~Ginges and V.~V.~Flambaum, 
Phys.\ Rep.\  {\bf 397}, 63 (2004);
%
M.~Pospelov and A.~Ritz, 
Annals Phys.\  {\bf 318}, 119 (2005);
%
T.~Fukuyama, 
Int.\ J.\ Mod.\ Phys.\ A {\bf 27}, 1230015 (2012);
%
J.~Engel, M.~J.~Ramsey-Musolf and U.~van Kolck, 
Prog.\ Part.\ Nucl.\ Phys.\  {\bf 71}, 21 (2013);
%
N. Yamanaka, 
{\it Analysis of the Electric Dipole Moment in the R-parity Violating Supersymmetric Standard Model}, (Springer, Berlin, 2014);
J.~de Vries and U.~G.~Mei{\ss}ner, 
Int.\ J.\ Mod.\ Phys.\ E {\bf 25}, 
1641008 (2016);
%
N.~Yamanaka, Int.\ J.\ Mod.\ Phys.\ E {\bf 26}, 
1730002 (2017);
%
N.~Yamanaka, B.~K.~Sahoo, N.~Yoshinaga, T.~Sato, K.~Asahi and B.~P.~Das, Eur.\ Phys.\ J.\ A {\bf 53}, 
54 (2017);
%
P.~Chang, K.~F.~Chen and W.~S.~Hou, Prog.\ Part.\ Nucl.\ Phys.\  {\bf 97}, 261 (2017);
%
T.~E.~Chupp, P.~Fierlinger, M.~J.~Ramsey-Musolf and J.~T.~Singh, Rev. Mod. Phys. {\bf 91}, 015001 (2019);
%
M.~S.~Safronova, D.~Budker, D.~DeMille, Derek~F.~Jackson~Kimball, A.~Derevianko and Charles~W.~Clark, Rev.\ Mod.\ Phys.\  {\bf 90}, 
025008 (2018);
%
N.~Yamanaka, PoS SPIN {\bf 2018}, 094 (2019) [arXiv:1902.00527 [hep-ph]].

\bibitem{Sakharov:1967dj}
A.~D.~Sakharov, Pisma Zh.\ Eksp.\ Teor.\ Fiz.\  {\bf 5}, 32 (1967) [JETP Lett.\  {\bf 5}, 24 (1967)] [Sov.\ Phys.\ Usp.\  {\bf 34}, 
392 (1991)] [Usp.\ Fiz.\ Nauk {\bf 161}, 
61 (1991)].

\bibitem{Farrar:1993hn}
G.~R.~Farrar and M.~E.~Shaposhnikov, Phys.\ Rev.\ D {\bf 50}, 774 (1994).

\bibitem{Huet:1994jb}
P.~Huet and E.~Sather, Phys.\ Rev.\ D {\bf 51}, 379 (1995).

\bibitem{Abel:2020gbr} 
C.~Abel {\it et al.} [nEDM Collaboration], Phys.\ Rev.\ Lett.\  {\bf 124}, 
081803 (2020).

\bibitem{Regan:2002ta}
B.~C.~Regan, E.~D.~Commins, C.~J.~Schmidt and D.~DeMille, Phys.\ Rev.\ Lett.\  {\bf 88}, 071805 (2002).

\bibitem{Bishof:2016uqx}
M.~Bishof {\it et al.}, Phys.\ Rev.\ C {\bf 94}, 
025501 (2016).

\bibitem{Graner:2016ses}
B.~Graner, Y.~Chen, E.~G.~Lindahl and B.~R.~Heckel, Phys.\ Rev.\ Lett.\  {\bf 116}, 
161601 (2016); {\bf 119}, 119901(E) (2017).

\bibitem{Sachdeva:2019mun} 
N.~Sachdeva {\it et al.}, 
Phys. Rev. Lett. {\bf 123}, 143003 (2019).

\bibitem{Hudson:2011zz}
J.~J.~Hudson, D.~M.~Kara, I.~J.~Smallman, B.~E.~Sauer, M.~R.~Tarbutt and E.~A.~Hinds, Nature {\bf 473}, 493 (2011).

\bibitem{Kara:2012ay}
D.~M.~Kara, I.~J.~Smallman, J.~J.~Hudson, B.~E.~Sauer, M.~R.~Tarbutt and E.~A.~Hinds, New J.\ Phys.\  {\bf 14}, 103051 (2012).

\bibitem{Baron:2013eja}
J.~Baron {\it et al.} [ACME Collaboration], Science {\bf 343}, 269 (2014).

\bibitem{Cairncross:2017fip}
W.~B.~Cairncross, D.~N.~Gresh, M.~Grau, K.~C.~Cossel, T.~S.~Roussy, Y.~Ni, Y.~Zhou, J.~Ye and E.~A.~Cornell, Phys.\ Rev.\ Lett.\  {\bf 119}, 
153001 (2017).

\bibitem{Andreev:2018ayy}
V.~Andreev {\it et al.} [ACME Collaboration], Nature {\bf 562}, no. 7727, 355 (2018).

\bibitem{Bennett:2008dy}
G.~W.~Bennett {\it et al.} [Muon (g-2) Collaboration], Phys.\ Rev.\ D {\bf 80}, 052008 (2009).

\bibitem{tauedm}
X.~Chen and Y.~Wu, 
J. High Energy Phys. 10 (2019) 089;
M.~K\"{o}ksal, A.~A.~Billur, A.~Guti\'{e}rrez-Rodr\'{i}guez and M.~A.~Hern\'{a}ndez-Ru\'{i}z, Phys.\ Rev.\ D {\bf 98}, 
015017 (2018);
M.~K\"{o}ksal, J. Phys. G {\bf 46}, 065003 (2019);
M.~K\"{o}ksal, A.~A.~Billur, A.~Guti\'{e}rrez-Rodr\'{i}guez and M.~A.~Hern\'{a}ndez-Ru\'{i}z, Int. J. Mod. Phys. A {\bf 34}, 1950076 (2019);
A.~Guti\'{e}rrez-Rodr\'{i}guez, M.~K\"{o}ksal, A.~A.~Billur and M.~A.~Hern\'{a}ndez-Ru\'{i}z, arXiv:1903.04135 [hep-ph];
L.~Beresford and J.~Liu, arXiv:1908.05180 [hep-ph].

\bibitem{EDMrelativistic}
J.~P.~Carrico, E.~Lipworth, P.~G.~H.~Sandars, T.~S.~Stein and M.~C.~Weisskopf, Phys.\ Rev.\  {\bf 174}, 125 (1968);
P. G. H. Sandars, Phys. Lett. {\bf 14}, 194 (1965);
{\bf 22}, 290 (1966);
V. V. Flambaum, Yad. Fiz. {\bf 24}, 383 (1976) [Sov. J. Nucl. Phys. {\bf 24}, 199 (1976)].

\bibitem{atomicedmtheory}
P.~G.~H.~Sandars and R.~M.~Sternheimer, Phys.\ Rev.\ A {\bf 11}, 473 (1975);
L. N. Labzovskii, Zh. Eksp. Teor. Fiz. {\bf 75}, 856 (1978) [Sov. Phys. JETP {\bf 48}, 434 (1978)];
O. P. Sushkov and V. V. Flambaum, Zh. Eksp. Teor. Fiz. {\bf 75}, 1208 (1978) [Sov. Phys. JETP {\bf 48}, 608 (1978)];
Z. W. Liu and H. P. Kelly, Phys. Rev. A {\bf 45}, R4210 (1992);
M. G. Kozlov and V. F. Ezhov, Phys. Rev. A {\bf 49}, 4502 (1994);
M.~G.~Kozlov and L.~N.~Labzowsky, J.\ Phys.\ B {\bf 28}, 1933 (1995);
T. M. R. Byrnes, V. A. Dzuba, V. V. Flambaum and D. W. Murray, Phys. Rev. A {\bf 59}, 3082 (1999);
M. K. Nayak and R. K. Chaudhuri, Chem. Phys. Lett. {\bf 419}, 191 (2006);
M. K. Nayak, R. K. Chaudhuri and B. P. Das, Phys. Rev. A {\bf 75}, 022510 (2007);
M. K. Nayak and R. K. Chaudhuri, Phys. Rev. A {\bf 78}, 012506 (2008);
H. S. Nataraj, B. K. Sahoo, B. P. Das, and D. Mukherjee, Phys. Rev. Lett. {\bf 101}, 033002 (2008);
D. Mukherjee, B. K. Sahoo, H. S. Nataraj, and B. P. Das, J. 
Phys. Chem. A {\bf 113}, 12549 (2009);
V. A. Dzuba and V. V. Flambaum, Phys. Rev. A {\bf 80}, 062509 (2009);
H.~S.~Nataraj, B.~K.~Sahoo, B.~P.~Das and D.~Mukherjee, Phys.\ Rev.\ Lett.\  {\bf 106}, 200403 (2011);
V. A. Dzuba, V. V. Flambaum and C. Harabati, Phys. Rev. A {\bf 84}, 052108 (2011); {\bf 85}, 029901(E) (2012);
S. G. Porsev, M. S. Safronova and M. G. Kozlov, Phys. Rev. Lett. {\bf 108}, 173001 (2012);
B. M. Roberts, V. A. Dzuba and V. V. Flambaum, Phys. Rev. A {\bf 88}, 042507 (2013);
D. V. Chubukov and L. N. Labzowsky, Phys. Lett. A {\bf 378}, 2857 (2014);
M. Abe, G. Gopakumar, M. Hada, B. P. Das, H. Tatewaki, and D. Mukherjee, Phys. Rev. A {\bf 90}, 022501 (2014);
A. Sunaga, M. Abe, M. Hada, and B. P. Das, Phys. Rev. A {\bf 93}, 042507 (2016);
L. Rad\u{z}i\={u}t\'{e}, G. Gaigalas, P. J\"{o}nsson and J. Biero\'{n}, Phys. Rev. A {\bf 93}, 062508 (2016);
A.~Sunaga, V.~S.~Prasannaa, M.~Abe, M.~Hada and B.~P.~Das, Phys.\ Rev.\ A {\bf 98}, 042511 (2018);
A.~Sunaga, M.~Abe, M.~Hada, and B.~P.~Das, Phys.\ Rev.\ A {\bf 99}, 062506 (2019);
M. Denis and T. Fleig, J. Chem. Phys. {\bf 145}, 
214307, (2016);
L. V. Skripnikov, J. Chem. Phys. {\bf 145}, 214301 (2016);
K. Talukdar, M. K. Nayak, N. Vaval, and S. Pal, J. Chem. Phys. {\bf 150}, 084304 (2019);
M.~Denis, P.~A.~B.~Haase, R.~G.~E.~Timmermans, E.~Eliav, N.~R.~Hutzler and A.~Borschevsky, Phys. Rev. A {\bf 99}, 042512 (2019);
N.~M.~Fazil, V.~S.~Prasannaa, K.~V.~P.~Latha, M.~Abe and B.~P.~Das, Phys.\ Rev.\ A {\bf 99}, 052502 (2019);
N.~Shitara, B.~K.~Sahoo, T.~Watanabe and B.~P.~Das, arXiv:1912.02981 [physics.atom-ph].

\bibitem{Sandars:1964zz} 
P.~G.~H.~Sandars and E.~Lipworth, Phys.\ Rev.\ Lett.\  {\bf 13}, 718 (1964).

\bibitem{Weisskopf:1968zz} 
M.~C.~Weisskopf, J.~P.~Carrico, H.~Gould and E.~Lipworth, Phys.\ Rev.\ Lett.\  {\bf 21}, 1645 (1968).

\bibitem{Stein:1969zz} 
T.~S.~Stein, J.~P.~Carrico, E.~Lipworth and M.~C.~Weisskopf, Phys.\ Rev.\  {\bf 186}, 39 (1969).

\bibitem{Player:1970zz}
M. A. Player and P. G. H. Sandars, J. Phys. B {\bf 3}, 1620 (1970).

\bibitem{Murthy:1989zz} 
S.~A.~Murthy, D.~Krause, Z.~L.~Li and L.~R.~Hunter, Phys.\ Rev.\ Lett.\  {\bf 63}, 965 (1989).

\bibitem{Abdullah:1990nh} 
K.~Abdullah, C.~Carlberg, E.~D.~Commins, H.~Gould and S.~B.~Ross, Phys.\ Rev.\ Lett.\  {\bf 65}, 2347 (1990).

\bibitem{Commins:1994gv} 
E.~D.~Commins, S.~B.~Ross, D.~DeMille and B.~C.~Regan, Phys.\ Rev.\ A {\bf 50}, 2960 (1994).

\bibitem{Chin:2001zz}
C.~Chin, V.~Leiber, V.~Vuletic, A.~J.~Kerman and S.~Chu, Phys.\ Rev.\ A {\bf 63}, 033401 (2001).

\bibitem{Sakemi:2011zz}
Y.~Sakemi {\it et al.}, J.\ Phys.\ Conf.\ Ser.\  {\bf 302}, 012051 (2011).

\bibitem{Kozyryev:2017cwq} 
I.~Kozyryev and N.~R.~Hutzler, Phys.\ Rev.\ Lett.\  {\bf 119}, 133002 (2017).

\bibitem{Aggarwal:2018pru} 
P.~Aggarwal {\it et al.} [NL-eEDM Collaboration], Eur.\ Phys.\ J.\ D {\bf 72}, 197 (2018).

\bibitem{Vutha:2017pej}
A.~C.~Vutha, M.~Horbatsch and E.~A.~Hessels, Atoms {\bf 6}, 3 (2018).  


\bibitem{GREDM}
A.~Kobach, Nucl.\ Phys.\ B {\bf 911}, 206 (2016);
A.~J.~Silenko and O.~V.~Teryaev, Phys.\ Rev.\ D {\bf 76}, 061101(R) (2007);
Y.~N.~Obukhov, A.~J.~Silenko, and O.~V.~Teryaev, Phys.\ Rev.\ D {\bf 94}, 044019 (2016);
Int.\ J.\ Mod.\ Phys.\ A {\bf 31}, 1645030 (2016);
Y. Orlov, E. Flanagan, and Y. Semertzidis, Phys. Lett. A {\bf 376}, 2822 (2012);
A.~L\'{a}szl\'{o} and Z.~Zimbor\'{a}s, Class.\ Quant.\ Grav.\  {\bf 35}, 175003 (2018);
A.~Laszlo, PoS SPIN {\bf 2018}, 182 (2019) [arXiv:1901.06217 [gr-qc]].

\bibitem{Kobayashi:1973fv}
M.~Kobayashi and T.~Maskawa, Prog.\ Theor.\ Phys.\  {\bf 49}, 652 (1973).

\bibitem{Shabalin:1978rs}
E.~P.~Shabalin, Sov.\ J.\ Nucl.\ Phys.\  {\bf 28}, 75 (1978) [Yad.\ Fiz.\  {\bf 28}, 151 (1978)].

\bibitem{Shabalin:1980tf}
E.~P.~Shabalin, Yad.\ Fiz.\  {\bf 31}, 1665 (1980).

\bibitem{Shabalin:1982sg} 
E.~P.~Shabalin, Sov.\ Phys.\ Usp.\  {\bf 26}, 297 (1983) [Usp.\ Fiz.\ Nauk {\bf 139}, 561 (1983)].

\bibitem{Eeg:1982qm} 
J.~O.~Eeg and I.~Picek, Phys.\ Lett.\  {\bf 130B}, 308 (1983).

\bibitem{Eeg:1983mt} 
J.~O.~Eeg and I.~Picek, Nucl.\ Phys.\ B {\bf 244}, 77 (1984).

\bibitem{Khriplovich:1985jr}
I.~B.~Khriplovich, Phys.\ Lett.\ B {\bf 173}, 193 (1986) [Sov.\ J.\ Nucl.\ Phys.\  {\bf 44}, 659 (1986)] [Yad.\ Fiz.\  {\bf 44}, 1019 (1986)].

\bibitem{Czarnecki:1997bu}
A.~Czarnecki and B.~Krause, Phys.\ Rev.\ Lett.\  {\bf 78}, 4339 (1997).

\bibitem{Pospelov:1991zt}
M.~E.~Pospelov and I.~B.~Khriplovich, Sov.\ J.\ Nucl.\ Phys.\  {\bf 53}, 638 (1991) [Yad.\ Fiz.\  {\bf 53}, 1030 (1991)].

\bibitem{Booth:1993af}
M.~J.~Booth, arXiv:hep-ph/9301293.

\bibitem{Pospelov:2013sca}
M.~Pospelov and A.~Ritz, Phys.\ Rev.\ D {\bf 89}, 056006 (2014).

\bibitem{Pospelov:1994uf}
M.~E.~Pospelov, Phys.\ Lett.\ B {\bf 328}, 441 (1994).

\bibitem{Jarlskog:1985ht}
C.~Jarlskog, Phys.\ Rev.\ Lett.\  {\bf 55}, 1039 (1985).

\bibitem{Glashow:1970gm}
S.~L.~Glashow, J.~Iliopoulos and L.~Maiani, Phys.\ Rev.\ D {\bf 2}, 1285 (1970).

\bibitem{Ellis:1976fn} 
J.~R.~Ellis, M.~K.~Gaillard and D.~V.~Nanopoulos, Nucl.\ Phys.\ B {\bf 109}, 213 (1976).

\bibitem{Yamaguchi:2020dsy} 
Y.~Yamaguchi and N.~Yamanaka, Phys.\ Rev.\ D {\bf 103}, 013001  (2021).


\bibitem{Khriplovich:1981ca}
I.~B.~Khriplovich and A.~R.~Zhitnitsky, Phys.\ Lett.\  {\bf 109B}, 490 (1982).

\bibitem{McKellar:1987tf}
B.~H.~J.~McKellar, S.~R.~Choudhury, X.~G.~He and S.~Pakvasa, Phys.\ Lett.\ B {\bf 197}, 556 (1987).

\bibitem{Seng:2014lea}
C.~Y.~Seng, Phys.\ Rev.\ C {\bf 91}, 025502 (2015).

\bibitem{Yamanaka:2015ncb}
N.~Yamanaka and E.~Hiyama, 
J. High Energy Phys. 02 (2016) 067.

\bibitem{Yamanaka:2016fjj}
N.~Yamanaka, Nucl.\ Phys.\ A {\bf 963}, 33 (2017).

\bibitem{Lee:2018flm} 
J.~Lee, N.~Yamanaka, and E.~Hiyama, Phys. Rev. C {\bf 99}, 055503 (2019).

\bibitem{HLS}
M.~Bando, T.~Kugo, S.~Uehara, K.~Yamawaki and T.~Yanagida, Phys.\ Rev.\ Lett.\  {\bf 54}, 1215 (1985);
M.~Bando, T.~Kugo and K.~Yamawaki, Prog.\ Theor.\ Phys.\  {\bf 73}, 1541 (1985);
Nucl.\ Phys.\ B {\bf 259}, 493 (1985);
M.~Bando, T.~Kugo and K.~Yamawaki, Phys.\ Rept.\  {\bf 164}, 217 (1988);
U.~G.~Meissner, Phys.\ Rept.\  {\bf 161}, 213 (1988);
F.~Klingl, N.~Kaiser and W.~Weise, Z.\ Phys.\ A {\bf 356}, 193 (1996);
M.~Harada and K.~Yamawaki, Phys.\ Rev.\ Lett.\  {\bf 86}, 757 (2001);
M.~Harada and K.~Yamawaki, Phys.\ Rept.\  {\bf 381}, 1 (2003).

\bibitem{Buchalla:1995vs}
G.~Buchalla, A.~J.~Buras and M.~E.~Lautenbacher,
Rev.\ Mod.\ Phys.\ \textbf{68} (1996), 1125-1144.

\bibitem{decayconstant}
M.~Neubert and B.~Stech, Adv.\ Ser.\ Direct.\ High Energy Phys.\  {\bf 15}, 294 (1998);
Y.~Grossman, M.~K\"onig and M.~Neubert, 
J. High Energy Phys. 04 (2015) 101; 
A.~Bharucha, D.~M.~Straub and R.~Zwicky, 
J. High Energy Phys. 08 (2016) 098;
Q.~Chang, X.~N.~Li, X.~Q.~Li and F.~Su, Chin.\ Phys.\ C {\bf 42}, 
073102 (2018).
  
\bibitem{buras}
A.~J.~Buras, M.~Jamin, M.~E.~Lautenbacher and P.~H.~Weisz, Nucl.\ Phys.\ B {\bf 370}, 69 (1992) 
{\bf 375}, 501(A) (1992);
G.~Buchalla, A.~J.~Buras and M.~E.~Lautenbacher, Rev.\ Mod.\ Phys.\  {\bf 68}, 1125 (1996).

\bibitem{Lee:1972px} 
B.~W.~Lee, J.~R.~Primack and S.~B.~Treiman, Phys.\ Rev.\ D {\bf 7}, 510 (1973).

\bibitem{Shrock:1978dm} 
R.~E.~Shrock and S.~B.~Treiman, Phys.\ Rev.\ D {\bf 19}, 2148 (1979).

\bibitem{Leinweber:2001ac} 
D.~B.~Leinweber, A.~W.~Thomas, K.~Tsushima and S.~V.~Wright, Phys.\ Rev.\ D {\bf 64}, 094502 (2001).

\bibitem{Guo:2018zvl} 
X.~Y.~Guo and M.~F.~M.~Lutz, Nucl.\ Phys.\ A {\bf 988}, 48 (2019).


\bibitem{Blum:2019ugy}
T.~Blum, N.~Christ, M.~Hayakawa, T.~Izubuchi, L.~Jin, C.~Jung and C.~Lehner,
Phys. Rev. Lett. \textbf{124} 132002 (2020).

\bibitem{Bai:2015nea}
Z.~Bai \textit{et al.} [RBC and UKQCD Collaborations],
Phys. Rev. Lett. \textbf{115}, 212001 (2015).

\bibitem{PhysRevA.93.062503} 
 D.~V.~Chubukov and L.~N.~Labzowsky, 
 Phys.\ Rev.\ A {\bf 93}, 062503 (2016).

\bibitem{Hiller:2003js} 
G.~Hiller and F.~Kruger, Phys.\ Rev.\ D {\bf 69}, 074020 (2004).

\bibitem{Bfactory}
J.~P.~Lees {\it et al.} [BaBar Collaboration], Phys.\ Rev.\ Lett.\  {\bf 109}, 101802 (2012);
J.~P.~Lees {\it et al.} [BaBar Collaboration], Phys.\ Rev.\ D {\bf 88}, 072012 (2013);
R.~Aaij {\it et al.} [LHCb Collaboration], Phys.\ Rev.\ Lett.\  {\bf 113}, 151601 (2014);
{\bf 115}, 111803(E) (2015); {\bf 115}, 159901 (2015);
S.~Wehle {\it et al.} [Belle Collaboration], Phys.\ Rev.\ Lett.\  {\bf 118}, 111801 (2017);
R.~Aaij {\it et al.} [LHCb Collaboration], 
J. High Energy Phys. 08 (2017) 055;
S.~Hirose {\it et al.} [Belle Collaboration], Phys.\ Rev.\ Lett.\  {\bf 118}, 211801 (2017);
S.~Hirose {\it et al.} [Belle Collaboration], Phys.\ Rev.\ D {\bf 97}, 012004 (2018);
R.~Aaij {\it et al.} [LHCb Collaboration], Phys.\ Rev.\ Lett.\  {\bf 120}, 171802 (2018);
R.~Aaij {\it et al.} [LHCb Collaboration], Phys.\ Rev.\ D {\bf 97}, 072013 (2018);
R.~Aaij {\it et al.} [LHCb Collaboration], Phys.\ Rev.\ Lett.\  {\bf 122}, 191801 (2019).

\bibitem{KOTOresult}
S. Shinohara, 
{\it Search for the rare decay $K_L \to \pi^0 \nu \bar{\nu}$ at J-PARC KOTO experiment, 
The International Conference on Kaon Physics 2019 (KAON2019)
} (2019).


\bibitem{Kitahara:2019lws}
T.~Kitahara, T.~Okui, G.~Perez, Y.~Soreq and K.~Tobioka,
Phys. Rev. Lett. \textbf{124},
071801 (2020)

\end{thebibliography}
\end{document}